# A Structuration Approach to Theorizing Cybersecurity Practice: The STARC Model

**Md Aktaruzzaman, Atif Ahmad and Sean Maynard**
School of Computing and Information Systems, The University of Melbourne, Victoria, Australia
Corresponding Author: m.aktaruzzaman@unimelb.edu.au

## Abstract

- **The problem.** Cybersecurity practice runs simultaneously across analysts, teams, organizations, sectors, and regulators, co-evolves with adversaries, and increasingly blends human and algorithmic decision-making. The theories applied to it operate at single organizational levels and cannot explain why organizations with broadly similar controls differ sharply in resilience.
- **This paper.** We develop STARC (Structuration Theory Adaptation for Resilient Cybersecurity), a framework for locating where cybersecurity practice succeeds or fails structurally. It extends Giddens' Structuration Theory with three innovations, Multi-Level Adversarial Agency, Threat-Adaptive Structuration, and Material-Agential Structural Properties, across five structure-agency triads.
- **Evidence base.** STARC is illustrated through re-analysis of three financial organizations, an Australian, an Indonesian, and a Malaysian bank, across 20 interviews from SOC analysts to senior executives, selected as diverse insourced and outsourced configurations rather than as a comparison of equivalents.
- **Cybersecurity contribution.** STARC offers a structural account of why differently resourced and outsourced organizations differ in resilience, and a vocabulary for diagnosing incident-response breakdown across levels, tempos, and the human-algorithm authority boundary that single-level frameworks leave invisible.
- **Theory and outputs.** It extends Structuration Theory to adversarial, multi-level, and hybrid human-algorithmic contexts, and yields seven testable propositions linking structuration to resilience, offered for future testing.

## 1. Introduction

The landscape of cybersecurity practice is rapidly evolving, necessitating robust theoretical frameworks capable of enhancing our understanding of cybersecurity practices (Shillair et al. 2022). To date, Situational Awareness Theory, Organizational Learning Theory, Strategy As Practice, Neo-Institutional Theory, and Sensemaking Theory have been predominant frameworks applied to cybersecurity practice (Ahmad et al. 2021; Ahmad et al. 2015). However, despite their contributions, these theories provide limited insight into the multifaceted nature of cybersecurity operations and the dynamic relationships between individual actions and organizational structures (Folorunsho et al. 2019; Prümmer et al. 2024). The scale of this challenge is substantial: the FBI (2023) reported losses exceeding US$10.3 billion from cybercrime in 2022 alone, while ACCC (2023) documented over AU$3.1 billion in scam losses in Australia, underscoring the practical urgency of stronger theoretical foundations.

This paper addresses these theoretical gaps by developing the Structuration Theory Adaptation for Resilient Cybersecurity (STARC) model. STARC introduces three innovations. First, Multi-Level Adversarial Agency (MLAA) theorizes cybersecurity decisions as operating simultaneously across individual, team, organizational, sector, and institutional levels, with each level recursively shaping the others - a cross-level recursive dynamic that single-level theories do not explicitly foreground. Second, Threat-Adaptive Structuration (TAS) explains how organizations maintain stable governance frameworks while adapting responses to emerging threats, introducing the concepts of bounded mutability and multi-temporal structuration. Third, Material-Agential Structural Properties (MASP) recognizes that security technologies exercise forms of agency through automated detection and response across a spectrum from rule-based execution to machine-learning-driven autonomy, extending Structuration Theory beyond its anthropocentric foundations to theorize hybrid human-algorithmic decision-making.

These three innovations are integrated in STARC's five-triad framework - routine decision-making, communication, power dynamics, policy and compliance, and technology and human resources - each adapted from Giddens' (1984) original structuration triads to map cybersecurity practice. Unlike earlier Structuration Theory applications in information systems that focused on technology adoption in primarily cooperative settings (DeSanctis & Poole 1994; Orlikowski 2000), STARC theorizes how structures and agency interplay across organizational levels in adversarial environments.

This study addresses the research question: *'How can the application of Structuration Theory improve our understanding of cybersecurity practice?'* Illustrative evidence is drawn from detailed investigation of cybersecurity practices across three financial organizations operating in Australia, Indonesia, and Malaysia.

This research makes three primary contributions. First, it extends Structuration Theory by theorizing multi-level agency in adversarial contexts, bounded mutability and multi-temporal structuration, and hybrid human-algorithmic agency. Second, it offers a unifying meta-theoretical framework integrating disparate cybersecurity theories by showing how their explanations recursively interact across levels. Third, it advances cybersecurity theory by proposing seven testable propositions linking structuration processes to resilience outcomes.

**Practical Value.** Beyond its theoretical contributions, STARC offers practical value for cybersecurity practitioners. The five-triad framework provides diagnostic tools for assessing security posture across operational, communicative, authority, compliance, and resource dimensions simultaneously. Bounded mutability gives practitioners a design principle for deliberately maintaining stable governance while enabling rapid peripheral adaptation to emerging threats, and MASP addresses the urgent governance challenge of AI adoption by providing apparatus for managing the boundary between algorithmic decision-making and human oversight. Section 6.3 develops these implications in detail.

Section 2 reviews existing cybersecurity theories and Structuration Theory foundations. Section 3 outlines the research design. Section 4 presents STARC's three theoretical extensions, five-triad framework, and testable propositions. Section 5 demonstrates analytical utility through theoretical comparison with prior analyses. Sections 6 and 7 discuss contributions, limitations, and future research.

## 2. Literature Review

This literature review examines how theory is used in cybersecurity research and why existing theories and frameworks provide only partial explanations of the dynamics STARC addresses.

### 2.1 The Role of Theory in Cybersecurity Research

Theory serves purposes beyond description - it provides explanation, prediction, and prescriptive guidance (Gregor 2006). In cybersecurity research, proliferation of case studies, best practice frameworks, and technical solutions has not been matched by commensurate theoretical development (Folorunsho et al. 2019). Good cybersecurity theory should explain why certain practices produce resilience while others fail, predict how changes in one element affect overall security posture, identify causal mechanisms linking actions to outcomes, and prescribe interventions grounded in understanding of underlying dynamics.

Cybersecurity exhibits four characteristics that create unique theoretical demands: *adversarial dynamics* (intelligent adversaries who probe structural weaknesses and adapt to defenses); *multi-level decisions* (spanning analyst triage, team coordination, organizational policy, sector intelligence sharing, and regulatory compliance, recursively interactive rather than hierarchically nested); a *temporal paradox* (simultaneous monthly-to-yearly governance stability and hourly-to-daily tactical adaptability); and *socio-technical complexity* (automated systems exercising algorithmic agency, creating hybrid decision-making that transcends anthropocentric assumptions).

## 2.2 Existing Theories and Frameworks: Strengths and Scope Limitations

This section systematically reviews theories applied to cybersecurity, identifying contributions and scope limitations. Table 1 provides a comparative overview. The intent is not to dismiss prior theories as inadequate - each addresses important aspects of cybersecurity practice - but to identify the specific analytical gaps that motivate STARC's integrative extensions.

Situational Awareness Theory [SAT] (Endsley 1995) explains how individuals perceive, comprehend, and project environmental elements (Ahmad et al. 2021). Its focus on immediate tactical awareness, however, provides limited apparatus for the recursive connections between individual awareness and the organizational and sector-level structures that shape it (Endsley 2015; Flach 1995). Moreover, rapid cyber threats can overwhelm human cognition (Dekker 2015; Parasuraman et al. 2008).

Organizational Learning Theory [OLT], particularly the 4I model (Zietsma et al. 2002), explains how learning progresses from individual intuition to institutionalized knowledge (Ahmad et al. 2019; Ahmad et al. 2015). However, its staged learning model does not explicitly theorize the simultaneous stability and rapid adaptation that effective cybersecurity demands (Milway and Saxton 2011).

Strategy as Practice [SAP] (Golsorkhi et al. 2015; Vaara and Whittington 2012) highlights concrete actions professionals take to protect digital assets. However, its emphasis on micro-level activities can produce siloed efforts and weakened strategic coherence (Jarzabkowski et al. 2007; Whittington 2015), and does not integrate cross-level adversarial dynamics.

Neo-Institutional Theory (Alvesson and Spicer 2019) examines how institutional structures shape organizational behavior, clarifying how regulations influence practices. However, it provides insufficient apparatus for individual agency and cross-border cyber threats (Patterson et al. 2024).

Sensemaking Theory (Weick et al. 2005) explains how individuals and organizations interpret complex, ambiguous situations. However, its emphasis on individual cognitive processes does not fully capture broader organizational structures shaping sensemaking (Combe and Carrington 2015).

Protection Motivation Theory (Rogers 1975; Boss et al. 2015) explains security behavior through threat and coping appraisal. As an individual-level cognitive theory, it does not address organizational structures, collective action, or the interplay of individual motivations with institutional constraints.

NIST Cybersecurity Framework 2.0 (Pascoe et al. 2024) offers structured prescriptive guidance through six core functions. Its primary limitation is that it clarifies *what* organizations should do but provides little insight into *why* practices produce resilience or *how* structures and actions interact. Similarly, ISO/IEC 27001 provides compliance requirements without explaining how controls function in practice or adapt to evolving threats.

*Table 1: Comparative Analysis of Existing Theories*

| Theory | Level | Key Contribution | Scope Limitation | Gap STARC Addresses |
|---|---|---|---|---|
| Situational Awareness | Individual | Tactical awareness in dynamic environments | Provides limited conceptual apparatus for cross-level recursive dynamics | MLAA |
| Organizational Learning | Organizational | How organizations learn from incidents | Assumes linear learning; does not explicitly theorize temporal differentiation | TAS |
| Strategy as Practice | Organizational | Day-to-day strategic activities | Addresses micro-level activities without integrating cross-level dynamics | MLAA + TAS |
| Neo-Institutional | Institutional | Regulatory/normative pressures | Provides partial but insufficient apparatus for individual agency and global dynamics | MLAA + MASP |

| Theory | Level | Key Contribution | Scope Limitation | Gap STARC Addresses |
|---|---|---|---|---|
| Sensemaking | Individual/Team | Interpreting ambiguous situations | Does not extend to the cross-level recursive processes STARC addresses | MLAA |
| Protection Motivation | Individual | Security behavior motivation | Was not designed to address collective action or institutional constraints | MLAA |
| NIST CSF 2.0 | Prescriptive | What to do | Taxonomy rather than explanatory theory; does not explain why practices produce resilience | All extensions |
| ISO 27001 | Prescriptive | Compliance requirements | Does not explain how controls function in practice or adapt to evolving threats | All extensions |

Taken together, these scope limitations share a common thread: existing theories were developed to explain stable, single-level, predominantly human phenomena, and each addresses only part of the recursive, multi-level, adversarial, and socio-technical character of cybersecurity practice identified in Section 2.1. An integrative foundation is therefore needed - one capable of theorizing structure and agency as mutually constitutive across levels, over time, and across the human-algorithmic boundary, rather than treating these as separate add-ons to a single-level theory. Structuration Theory is well suited to this role precisely because it was developed as a general theory of the structure-agency relationship rather than a domain-specific account, giving it the conceptual flexibility to be extended - through MLAA, TAS, and MASP - to address the gaps identified above without requiring a wholesale alternative framework.

### 2.3 Structuration Theory Foundations

Structuration Theory addresses a fundamental sociological question: what constitutes society? Giddens (1984) proposes the concept of 'duality of structure,' where structure simultaneously shapes actions (as medium) and is shaped by those actions (as outcome). This challenges traditional divisions between subjective individual-micro and objective society-macro perspectives (Andersson 2010). Structure and agency are so interdependent that comprehending one requires considering the other (Sewell 1992).

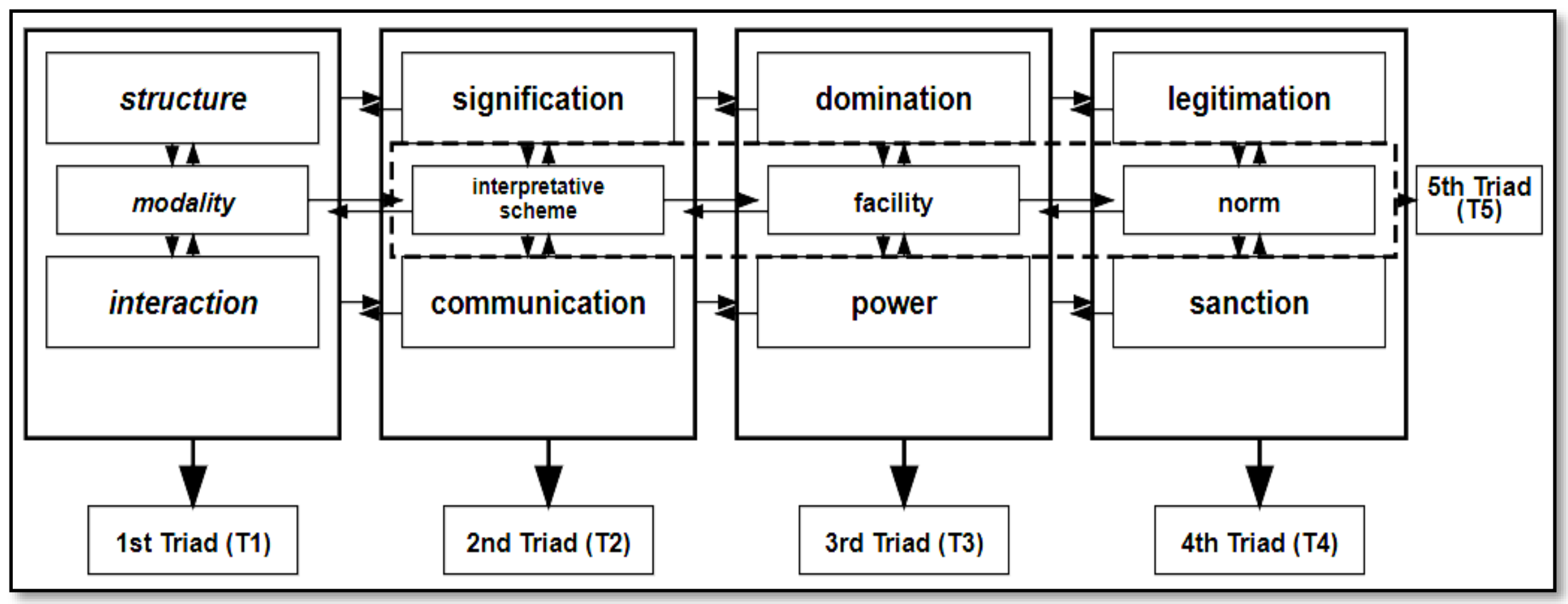

*Figure 1. Perspectives of the duality of structure (Source: Giddens 1984, p. 29)*

Structuration Theory consists of three interconnected layers: structure (structural properties of social systems), agency (interaction), and modalities (mechanisms translating structure into action). Giddens (1984) outlines three structural dimensions - signification, domination, and legitimation - aligned with corresponding agency dimensions: communication, power, and sanction, mediated through interpretative schemes, facilities, and norms. As illustrated in Figure 1, the theory can be understood through five interconnected triads: (T1) Structure-Modality-Interaction; (T2) Signification-Interpretative Scheme-Communication; (T3) Domination-Facility-Power; (T4) Legitimation-Norm-Sanction; and (T5) Interpretative Scheme-Facility-Norm. These five triads form meaningful analytical building blocks upon which the STARC model is constructed.

### 2.4 ST Evolution in Information Systems and Beyond

Structuration Theory has profoundly shaped Information Systems (Jones and Karsten 2008). Orlikowski (1992, 2000) established foundational concepts through the 'duality of technology' and 'technology-in-practice.' DeSanctis and Poole's (1994) Adaptive Structuration Theory (AST) extended this by examining organizational change through structures embedded in technologies and structures emerging through human-technology interactions; Poole and DeSanctis (2006) further developed micro-level structuration dynamics. DeSanctis and Gallupe (1987) provided early foundations for understanding group decision support in technology-mediated settings that informed AST's development.

More recent scholarships have developed increasingly nuanced understandings. Leonardi (2011, 2013) introduced sociomateriality theory, arguing that while social and material are inseparable in practice, they must be analytically distinguished - providing a theoretical foundation for MASP. Volkoff and Strong (2013) reinforced this through critical realist foundations, demonstrating that security controls have real material effects regardless of user perception. Baptista et al. (2020) showed how technology platforms create generative structures enabling continuous innovation through recursive appropriation, directly informing TAS's theorization of continuous defensive adaptation.

Stones' (2005) strong structuration extends Giddens by attending more explicitly to agents' conjunctural knowledge and the position-practice system - directly relevant to MLAA's theorization of how agents at different levels draw on position-specific knowledge in cybersecurity decisions. Actor-Network Theory (Latour 2005; Callon 1986) offers an alternative perspective on non-human agency, theorizing non-human actants without reproducing the structure/agency duality central to ST; Prashantham and Healey (2022) inform the strategic practice dimensions of STARC. STARC does not import ANT's actant concept, which would be inconsistent with its rejection of the structure/agency binary, but instead extends ST's own apparatus to accommodate algorithmic structural properties absent from Giddens' original formulation. Unlike sociomateriality theory, which treats social and material as analytically inseparable, MASP maintains analytical distinction between human and algorithmic agency precisely because the governance challenge - deciding which authority domain applies - requires that distinction.

### 2.5 Limitations of ST Applications in Cybersecurity

Despite rich theoretical development, existing ST applications exhibit four gaps when applied to cybersecurity. First, they primarily theorize agency at single organizational levels, providing inadequate apparatus for the simultaneous multi-level agency that cybersecurity demands (Folorunsho et al. 2019; Ibrahim et al. 2024). Unlike multi-level analysis in organizational research - which provides methodological tools for statistical aggregation across levels - MLAA theorizes the recursive mechanisms through which agency at one level constitutes and is constituted by agency at other levels within an adversarial context. Second, ST applications predominantly assume cooperative or neutral technology contexts, lacking conceptual apparatus for external actors intentionally probing structural weaknesses. Third, existing applications assume structural durability with gradual evolution (Baptista et al. 2020), whereas cybersecurity requires structures exhibiting simultaneous stability and rapid adaptability. Unlike organizational ambidexterity theory, which identifies the need for simultaneous stability and adaptation without specifying the structural mechanism, TAS theorizes bounded mutability as the specific structuration process through which this is achieved, and adversarial intelligence as the exogenous driver that distinguishes this from ambidexterity in cooperative contexts. Fourth, while recent work acknowledges material agency (Leonardi 2013; Volkoff and Strong 2013), it remains focused on how material properties constrain human agency rather than theorizing autonomous algorithmic agency requiring meta-governance, addressed through MASP.

## 3. Research Design and Methodology

### 3.1 Theoretical Development Research Approach

This research represents theoretical development rather than primary empirical hypothesis testing. Following established IS theory development traditions (Gregor 2006; Weber 2012), it proceeds through five stages: identifying theoretical gaps through literature review; developing theoretical extensions addressing those gaps; constructing an integrated model; illustrative application demonstrating analytical utility; and generating testable propositions for future research. This approach has precedent in IS theory development: Orlikowski's (2000) Technology-in-Practice and Leonardi's (2011) Imbrication Theory both advanced understanding through theoretical argumentation complemented by empirical illustration.

### 3.2 Three-Case Re-Analysis as Theoretical Illustration

To illustrate STARC's utility, this study re-analyzes existing qualitative datasets from three financial organizations - an Australian, an Indonesian, and a Malaysian bank. None of these datasets was collected to test STARC; all were gathered in prior cybersecurity-practice research, revisited here through the STARC lens. The Australian dataset, on situational awareness and incident response, was examined previously through Situational Awareness Theory (Ahmad et al. 2021) and Organizational Learning Theory (Kotsias et al. 2022); the Indonesian and Malaysian datasets came from parallel studies of cybersecurity governance in Southeast Asian banks. Re-analyzing pre-existing material rather than collecting data to fit the model guards against confirmatory evidence, a strength noted in theory-development research (Gregor 2006; c 2018).

We revisit these datasets for four reasons. First, financial organizations are theoretically appropriate sites: their resourcing and specialization - dedicated SOCs, specialist teams, sector-level intelligence sharing - let them exhibit the multi-level, multi-temporal, and socio-technical dynamics the model theorizes. Second, the datasets comprise 20 interviews across the full hierarchy, from Level 1 SOC analysts to CISOs and senior executives, letting recursive relationships among individual, team, organizational, sector, and institutional agency be examined as no single-level dataset could. Third, because the Australian material was already analyzed through Situational Awareness Theory and organizational learning, applying STARC foregrounds what it adds against a familiar baseline - an illustration of the model's reach, not a controlled test. Fourth, the cases differ deliberately in resourcing and outsourcing structure - the axis along which practice varies most in this sector - and are treated as diverse instantiations, not comparable points on a common scale, so their differences become analytical leverage rather than a confound.

The three-case design is theoretically motivated, not additive: the cases are not commensurable, so each follows a logic of *diverse instantiation* rather than ranking. The Australian bank - fully insourced, with a 24/7 SOC, automated end-to-end detection and response, and a developed sector community - renders the framework's mechanisms in *realized form*: bounded mutability, multi-level alignment, and an explicitly managed algorithmic-human boundary as they appear once fully instantiated. The Indonesian bank, which rebuilt its architecture after the 2018/2019 ransomware incident that exposed near-absent governance, renders structuration *in motion* - how core boundaries are constructed rather than how they appear once settled. The Malaysian bank, heavily reliant on a managed security service provider, with playbooks in draft and a CISO function created only in 2019 under Bank Negara's Risk Management in Technology (RMIT) mandate, renders the *outsourcing boundary* observable: where operational agency sits when triage is external, and what a governance core looks like before a cybersecurity-specific one exists. Read together, the three cases do not form a maturity ranking; they instantiate the framework across insourced and outsourced, stable and reconstructing, resourced and under-resourced configurations - a range a single case could not provide.

The procedure ran in two stages. First, transcripts were open-coded (Patton 2015) for recurring patterns in decision-making, information flows, authority structures, and technology use, without imposing STARC categories. Second, STARC concepts were applied deductively, mapping coded segments against the three extensions (MLAA, TAS, MASP) and five triads (T1–T5) in a structured NVivo framework with identical node hierarchies across the cases. Segments that resisted categorization or were equally well explained by prior frameworks were retained to refine the specification rather than discarded - for example, the Australian bank's post-incident learning gap at Level 1 and Indonesia's ambiguous decision authority are reported as negative or partial cases rather than suppressed. Representative quotes were selected for clarity, each source drawn on for distinct points to avoid over-reliance on any single moment. Re-analysis of three organizations in one sector cannot establish generalizability; the framework is therefore presented as a candidate for future testing through the designs in Table 2. The procedure mitigates but cannot eliminate confirmation bias, since deductive mapping used knowledge of the STARC categories and the proportion of segments resisting categorization is not separately reported - a common limitation of theory-development research, noted here explicitly.

## 4. The STARC Model: Theoretical Extensions and Five-Triad Framework

### 4.1 Extension 1: Multi-Level Adversarial Agency (MLAA)

The analytical value of MLAA is illustrated by a recurring pattern across all three cases: sector-level threat intelligence triggers organizational control deployment before attacks occur - a phenomenon that existing single-level theories do not explicitly foreground. The Australian bank Threat Intelligence Manager describes this proactive structuration:

> *"we deploy the controls proactively even before the attack happens, we are running that whole loop and deploying our controls because we know the potential actor is operating in our geography based on the intel." [Threat Intelligence Manager, Australian Bank]*

This represents sector-level intelligence (from Australia's closely-knit financial sector intelligence community, incorporating ACSC and peer banks) shaping organizational controls (threat intelligence team deploys detection rules) which modify individual analyst practices - all before any attack materializes. The same pattern appears in Indonesia, where the CIO describes sector intelligence enabling proactive defense:

> *"from my CSO... a peer bank has been hit, so I knew... we need to check our logs for intrusion, possible similar intrusions." [CIO, Indonesian Bank]*

In Malaysia, sector agency flows through FinTIP and informal inter-bank channels:

> *"There are a system initiated by Bank Negara. They call it FinTIP... where all the banks will share the incidents... based on those, my team will do the threat hunting." [CISO, Malaysian Bank]*

The cross-case comparison illuminates variation in how sector-level agency is configured across the three cases. The Australian bank participates in a highly developed sector community, built on personal relationships and sustained leadership investment, with multiple channels (Slack communities, bi-monthly conferences, shared TI platforms). The Threat Intelligence Manager notes: *"we have a very close group of threat intel community in Australia... this threat intelligence sharing through people in your region, in your sector, who you know, it's probably critical for us."* Indonesia has multiple overlapping communities (Perbanas, Himbara, BSSN), while Malaysia relies primarily on FinTIP supplemented by informal inter-bank contact.

Giddens (1984) conceptualizes agency primarily at individual and organizational levels, with limited attention to cross-level dynamics. MLAA addresses this gap by theorizing agency as recursively operating across five interactive levels:

- Individual Agency (Analyst Level): SOC analysts, threat intelligence specialists, and security engineers exercise knowledgeable agency in interpreting alerts, triaging incidents, and implementing controls.
- Team Agency (Operational Level): Cybersecurity teams collectively constitute practices through shared interpretative schemes and coordinated incident responses.
- Organizational Agency (Strategic Level): CISOs and security leadership exercise strategic agency in resource allocation, policy formulation, and risk acceptance decisions.
- Sector Agency (Collaborative Level): Industry-specific information sharing arrangements (e.g., financial sector ISACs, FinTIP) constitute collective agency through threat intelligence exchange.

- Institutional Agency (Regulatory Level): Regulatory bodies (APRA/ASIC in Australia, OJK/Bank Indonesia in Indonesia, Bank Negara in Malaysia) exercise meta-agency by establishing legitimation frameworks.

**Critical Innovation:** These levels are not hierarchically nested but recursively interactive. Unlike multi-level analysis in organizational research - which provides methodological tools for statistical aggregation across levels - MLAA theorizes the recursive mechanisms through which agency at one level constitutes and is constituted by agency at other levels within an adversarial context. Adversarial agency - the intentional, adaptive actions of threat actors - creates unique structuration dynamics not present in traditional ST applications.

**Empirical Illustration - Cross-Case Comparison:** The Australian bank instantiates five-level MLAA alignment in fully realized form. Its CISO describes organizational-level agency as bounded coordination: *"the accountability ultimately rests on the Security executive. The role therefore becomes one of coordination, a decision hub with thresholds of delegated rights."* This coordination is enabled by a recent restructuring initiative that removed middle-management layers so security leadership can reach domain leads directly. Indonesia's CISO operates a hub-and-spoke model with governance flowing both upward and laterally across the C-suite: *"My directors, risk management directors, consumer directors, all of them... They have my number. They can call me anytime."* Malaysia's CISO illustrates institutional agency reshaping organizational structure: *"In 2019, when Bank Negara issued the policy document on risk management and technology, there was a requirement for banks to have a CISO function... I was appointed to spearhead that initiative."*

Institutional agency operates differently across the three environments. The Australian bank's IR Team Leader describes institutional pressure shaping formal risk acceptance: *"it might be a risk that we have to accept for a period of time... that will go through the appropriate risk framework to be endorsed or accepted by the right people."* Indonesia faces a heavier burden, reconciling Bank Indonesia, NIST, and BSSN frameworks alongside a 2022 data-privacy law introducing criminal liability: *"if we don't do a good job at this then we risk jail time and penalties."* Adversarial agency also varies: the Australian bank's Threat Intelligence Manager notes a geographical lag affording advance warning - *"there's a lag between the tactics, techniques and procedures... in North America, Europe and then Australia. So, you have advance warning, in a sense, of what's coming"* - while Indonesia confronts adversaries recruiting employees (*"these threat actors actually actively contact our employees"*) and Malaysia's Head of Surveillance acknowledges an attribution gap (*"we do not have the capability yet to immediately identify that this is by APT"*).

**Theoretical Proposition 1 (P1):** Cyber resilience emerges from the recursive alignment of agency across multiple organizational levels, where misalignment between individual, organizational, sector, and institutional agency creates exploitable vulnerabilities.

### 4.2 Extension 2: Threat-Adaptive Structuration (TAS)

Tactical detection rules at the Australian bank are updated within hours in response to emerging vulnerabilities, while fundamental governance structures remain stable over years. The IR Team Leader describes rapid peripheral adaptation triggered by sector intelligence:

> *"What we did specific to both of those incidents, both of those vulnerabilities, is we were able to reverse engineer the exposure. We then built specific new detection rules both on our Web application firewalls for our Internet facing services but also on things like our IPS/IDS, intrusion prevention systems, and therefore lifted our barriers, strengthened our barriers to protect our environment." [IR Team Leader, Australian Bank]*

This peripheral adaptation (new WAF rules, IPS/IDS signatures, and detection logic deployed in response to sector intelligence on the WannaCry and Apache Struts vulnerabilities) occurred within hours to days, while core governance remained stable: the CSIRT charter pre-authorized such responses within defined categories, and the restructure provided a stable authority framework for cross-domain coordination without governance renegotiation.

In Indonesia, the most illuminating TAS finding comes from the contrast between the pre-2018 and post-2018 states. Pre-2018, the organization's CIO describes: *"a lot of governance were non-existed prior to that incident... people were doing things they know best... basically, it's a scramble mode."* The ransomware incident triggered years-long core structural reconstruction: establishing a 24/7 SOC, centralizing logging across two data centers, implementing SDLC security gates, and formalizing the CSIRT decree. Today, the Indonesian bank operates with explicit fast-cycle metrics: *"The mean time to detect is fifteen minutes... from layer one to layer two, it is fifteen minutes, a total of thirty minutes."* In Malaysia, the BCM color-code system (Amber/Red/Black) provides a core structural escalation scaffold, but playbooks remain in draft - the organization's core structural mechanism here is a general business-continuity ladder overseen by a largely non-cybersecurity committee that can take months to decide, distinct from a cybersecurity-specific core.

Existing theories are insufficient to resolve the stability-adaptability paradox: OLT assumes staged uniform learning; SAT addresses real-time cognition without theorizing enduring structural stability; SAP describes micro-level activities without explaining how stable strategic frameworks coexist with fast-evolving tactics; and Neo-Institutional Theory predicts inertia rather than bounded adaptation. Resolving the paradox requires theoretical apparatus explaining simultaneous stability and adaptability through temporal differentiation - Threat-Adaptive Structuration (TAS).

#### 4.2.1 Element-Level Temporal Differentiation: Bounded Mutability

Within any given cybersecurity structure, core elements remain stable while peripheral elements adapt rapidly within those stable boundaries: this is bounded mutability.

- Core Structural Elements (Slow-changing): Fundamental security principles, organizational security culture, basic governance frameworks, foundational technical architectures, CSIRT charters, regulatory compliance mandates.
- Peripheral Structural Elements (Fast-changing): Specific detection rules and threat signatures, tactical response procedures and playbooks, tool configurations, operational workflows, and IOC-based blocking lists.

**Critical Innovation:** Unlike organizational ambidexterity theory, which identifies the need for stability and adaptation without specifying the mechanism, TAS theorizes bounded mutability as the specific process achieving this, with adversarial intelligence as the exogenous driver distinguishing it from ambidexterity in cooperative contexts: core elements provide stable boundaries within which peripheral elements adapt without renegotiating core principles.

**Cross-Case Empirical Illustration:** the Australian bank's crypto-miner incident provides the benchmark bounded mutability example for the dataset. At 09:00 on 1 July, a threat actor began activities on a Google Cloud Platform container; at 09:02 automated detection rules fired; at 09:05 the environment was contained; by 09:20 the environment was completely rebuilt and service restored - a 20-minute end-to-end automated response. The core architecture (detection rules, automation framework) remained stable; peripheral response executed entirely within pre-defined governance bounds. In Indonesia, the SDLC gate exemplifies a mature core structural constraint: *"If you don't go to that system, you will not be able to, it will not be put in production."* Daily detection rule updates by the dedicated detection engineering team represent peripheral adaptation operating continuously within this stable core. In Malaysia, the BCM's RTO threshold (the general business-continuity mechanism described in Section 4.2) provides the pre-specified escalation trigger: *"the system disruption did not resolve within... our timeframe we call RTO. We need to go to DR. So we activate code red"*. This escalation runs through the same non-cyber-specific committee, not a dedicated cybersecurity core.

#### 4.2.2 System-Level Temporal Differentiation: Multi-Temporal Structuration

Beyond element-level bounded mutability, TAS theorizes that different structural domains operate at systematically different temporal scales. Fast cycles (minutes–hours) encompass T1, T2, and T5-tactical. Moderate cycles (hours–days) involve major incident command activation (T3) and tactical tool deployments. Slow cycles (months–years) govern policy frameworks (T4), strategic power structures (T3), and major architecture decisions (T5).

**Critical Innovation:** Different structural domains serve different organizational functions requiring different temporal dynamics; if all changed at the same rate, either tactical responsiveness would suffer or strategic coherence would collapse. Multi-temporal structuration enables organizations to be simultaneously responsive and coherent.

**Cross-Case Empirical Illustration:** Fast-cycle response takes different forms across the cases. The Australian bank's detection is fully automated end-to-end; Indonesia operates against contractual detection and response targets; Malaysia's fast cycle culminates in a manual MBRC activation rather than an automated response, a different process rather than a slower version of the same one. These differences are consistent with P4. At the slow-cycle level, all three have multi-year structural investments (the Australian bank's restructure, Indonesia's 2018–2021 reconstruction, Malaysia's three-year CSSP following the 2019 RMIT mandate).

### 4.2.3 Integration: How Bounded Mutability and Multi-Temporal Structuration Enable TAS

Bounded mutability ensures coherence within each triad; multi-temporal structuration enables coordination across triads. Together they enable something absent from traditional ST: active structural maintenance under continuous adversarial pressure. Where standard ST treats structure reproduction as relatively passive, TAS theorizes cybersecurity structures as requiring continuous reconstitution against adaptive adversaries.

**Theoretical Proposition 2 (P2):** Organizational cyber resilience is positively associated with the capacity to maintain core structural stability while enabling rapid peripheral structural adaptation in response to threat intelligence.

**Theoretical Proposition 3 (P3):** Organizations with rigid structures (low peripheral adaptability) and those with fluid structures (low core stability) both exhibit reduced cyber resilience compared to organizations achieving bounded mutability.

**Theoretical Proposition 4 (P4):** Cyber resilience is enhanced when organizations successfully differentiate temporal scales across triads, enabling rapid tactical adaptation (T1-T2-T5 fast cycles) while maintaining strategic stability (T3-T4-T5 slow cycles).

### 4.3 Extension 3: Material-Agential Structural Properties (MASP)

Across all three cases, automated systems exercise forms of agency that cannot be reduced to human interpretation. At the Australian bank, automated behavioral profiling flags anomalous employee actions without human intervention, as the Cyber Strategy Lead describes:

> *"we profile every staff member... if all of a sudden, I upload data to Dropbox or to GitHub or something like that, then we'll pick that up as anomalous Behavior." [Cyber Strategy Lead, Australian Bank]*

In Indonesia, the SIEM/QRadar pipeline processes 3TB of logs daily:

> *"the rule we use, Rule detection, detects an anomaly. Then it will create a ticket, and then the ticket will be validated by layer one." [SOC Analyst, Indonesian Bank]*

In Malaysia, automated performance monitoring provided the first signal in a Cobalt Strike incident: *"We see performance monitoring on the server suddenly been high and it's obvious show that something is happening and there are information that going outside."*

Automated systems currently handle high-confidence, high-volume pattern matching effectively - blocking known-malicious IP addresses, quarantining files matching known malware signatures, and alerting on statistically anomalous access - while human analysts retain authority over low-confidence, high-stakes decisions: interpreting contradictory signals, assessing novel techniques without prior signatures, and making risk-acceptance decisions requiring contextual knowledge algorithms cannot access.

Existing theories are uniformly anthropocentric - SAT, OLT, SAP, and Neo-Institutional Theory all assume human cognition as the locus of agency. Even Technology-in-Practice (Orlikowski 2000) treats technology as human-appropriated facilities rather than autonomous agents. MASP extends ST's anthropocentric foundations by identifying three structural properties unique to cybersecurity technologies:

- Autonomous Agential Capacity: Security technologies (SIEM, IDS/IPS, automated response systems) exercise forms of agency through algorithmic decision-making, automated responses, and pattern recognition that operate independently of immediate human action.
- Material Structural Constraints: Technical controls create hard constraints (blocked connections, denied access, enforced encryption) that differ from social structural constraints in their immediacy and enforceability. Indonesia's SDLC gate and post-2018 USB lockdowns exemplify this; Malaysia's CISO notes: 'I don't touch any of the firewalls. My team don't have any access to the DLP's to the firewall.'
- Hybrid Interpretative Schemes: Human analysts and algorithmic systems maintain different but interacting interpretative schemes - human contextual judgment and machine pattern recognition. The Australian bank's SOC Leader describes this: 'How do I work with imperfect data? How do I work with data where there's weird contradictions? How do I troubleshoot those contradictions and figure out through experience?'

**Critical Innovation:** Unlike sociomateriality theory, which treats social and material as analytically inseparable in practice, MASP maintains analytical distinction between human and algorithmic agency precisely because the governance challenge - deciding which authority domain applies - requires that distinction. The boundary between algorithmic and human agency is the object of theoretical and practical interest, not merely a feature of practice.

**Cross-Case Comparison of Meta-Governance:** Across the three cases, MASP foregrounds three different meta-governance questions rather than three positions on a single clarity scale. The Australian bank shows the boundary in realized form: the CISO explicitly decides on service takedowns (human override of automated recommendations), incident managers coordinate T1-T5 execution, and severity thresholds trigger governance transitions. The SOC Leader describes: *"Our recommendation is, we actually think this environment should be taken off line... So, we might come up with that recommendation, but it would go up to CISO level, if not beyond, to get that decision yes-or-no."* Indonesia shows the same boundary before it has settled - decision authority acknowledged as still being constituted rather than exercised: *"For now, the decision maker is not yet clear... the CISO recommends... But the main decision maker usually goes back to the application owner, to the business."* Malaysia shows the boundary extending outward to the managed security service provider relationship, where BCM advises, MBRC decides, and IT executes, so the governing question becomes where authority sits when triage is externally performed. Read as instantiation rather than ranking, each configuration makes a different facet of P5 observable.

**Theoretical Proposition 5 (P5):** Effective cyber resilience requires meta-governance structures that manage the boundary between automated algorithmic agency and human override authority, with misalignment creating either excessive rigidity (over-automation) or reactive vulnerability (under-automation).

### 4.4 How the Extensions Integrate

The following discussion uses the notation A+B to indicate that extensions A and B are mutually constitutive - each enables and depends upon the other. This is not additive but recursive.

**MLAA+TAS:** Multi-level agency enables threat-adaptive structuration because agency at different levels operates at different temporal scales: individual analysts adapt peripheral structures rapidly (minutes/hours) while institutional agency adapts core structures slowly (months/years), and this differentiation enables bounded mutability as sector-level intelligence (MLAA) triggers fast-cycle peripheral adaptation (TAS) before threats materialize locally.

**TAS+MASP:** Threat-adaptive structuration is partly achieved through material-agential structural properties: automated systems implement control changes faster than human processes allow - as demonstrated by the Australian bank's 20-minute response - while core architectural constraints maintain stability. Financial sector ISACs in STIX/TAXII format enable participating organizations to automatically ingest IoC data (malicious IPs, domains, file hashes, behavioral signatures) directly into SIEM and threat intelligence platforms, triggering rule updates without human intermediation at each step.

**MASP+MLAA:** Material-agential structural properties distribute agency across human and non-human actors at multiple levels: automated threat detection extends individual analyst agency, while AI-driven threat intelligence platforms enable sector-level collective agency. The Australian bank's deployment of controls based on sector intelligence before attacks materialize represents MASP (automated deployment) enabling MLAA (sector intelligence reaching analyst practices).

### 4.5 The Five-Triad Integrated Framework

Figure 2 presents the integrated STARC model, showing how the three extensions and five triads operate together across organizational levels within an adversarial environment.

The STARC Model
External Adversarial Environment — threat actors continuously probe all levels
Multi-Level Adversarial Agency (MLAA)
recursive agency across levels
Institutional
regulators, standards (slow-cycle)
Sector
ISACs, collaboration
Organizational
CISOs, governance
Team
SOC, IR practices
Individual
analysts, real-time (fast-cycle)
Five Interconnected Triads (T1–T5)
each triad enacted across every MLAA level
T1 · Routine Decision-Making
structure — modality — interaction
triage, escalation, SOPs
T2 · Communication
signification — int. scheme — communication
intel sharing, sensemaking
T3 · Power Dynamics
domination — facility — power
authority, incident command
T4 · Policy & Compliance
legitimation — norm — sanction
regulation, audit
T5 · Technology & Resources
int. scheme — facility — norm
people + platforms (SIEM, IDS/IPS)
Threat-Adaptive Structuration (TAS)
Bounded mutability
Core — stable
principles, governance
Periphery — adaptive
detection rules, tactics
Multi-temporal structuration
Rapid
minutes–hours
Moderate
hours–weeks
Slow
months–years
P7: second-order learning loop
Material-Agential Structural Properties (MASP)
algorithmic agency + meta-governance
Algorithmic agency
rule-based → ML-driven automated detection & response
Material constraints
enforced controls: blocking, quarantine, encryption
Hybrid schemes
human judgment + machine pattern recognition
Meta-governance
governs the human–algorithmic authority boundary (P5)
Emergent Outcome: Cyber Resilience
arising from recursive co-evolution of MLAA, TAS & MASP across all triads
Legend
recursive duality (two-way)
mutual constitution across extensions
P7 second-order learning feedback

*Figure 2. Structuration Theory Adaptation for Resilient Cybersecurity (STARC)*

As Figure 2 illustrates, Multi-Level Adversarial Agency (MLAA, left) spans institutional, sector, organizational, team, and individual agency, with two-way arrows showing the recursive duality of structure. Material-Agential Structural Properties (MASP, right) comprise algorithmic agency, material constraints, and hybrid interpretative schemes under a meta-governance boundary (P5). The five triads (T1-T5, center) are each enacted across every MLAA level. Threat-Adaptive Structuration (TAS) appears as both bounded mutability (core/periphery) and multi-temporal structuration (rapid, moderate, slow). Open arrows denote the mutual constitution (the "+" in Section 4.4) of the three extensions; the dashed path traces the P7 learning loop. Cyber resilience is the emergent outcome.

### 4.5.1 Triad 1: Routine Decision-Making (Operational Foundation)

Routine decision-making forms the operational foundation of cybersecurity work, encompassing incident classification, escalation matrices, and standard operating procedures, translated into practice through SIEM platforms and incident-tracking tools.

**MASP in T1** is most clearly illustrated through algorithmic versus human triage. Financial sector SOCs receive thousands to millions of security events daily - a scale impossible for human analysis alone. In all three cases, SIEM platforms assign severity scores, correlate events, and generate alerts based on detection rules before any human involvement. At the Australian bank, 90% of email is blocked autonomously at the perimeter before human review. The Level 1 analyst describes: *"I handle phishing and malware incidents that are usually fairly low priority incidents... If I do encounter anything that's more serious, I escalate it to level two."* Indonesia's SOC processes approximately 8.6 million attacks per week with QRadar. Malaysia's outsourced SOC performs L1–L3 triage algorithmically. However, T1 also reveals important maturity differences: The Australian bank's L1 analyst describes an 18-step manual phishing process taking 20 minutes per ticket due to tool integration gaps - *"The phishing handling process that I currently have to follow is extremely manual"* - illustrating that sophisticated T5 infrastructure does not guarantee T1 efficiency at all analyst levels.

**MLAA and TAS in T1:** analysts exercise sector-influenced agency that recursively reshapes organizational standards, while stable incident categorization coexists with adaptive triage criteria - illustrated by Indonesia's kill-chain playbooks and the Australian bank's two-tier (procedural + problem-solving) playbooks developed with Mandiant.

### 4.5.2 Triad 2: Communication in Decision-Making (Information Flow)

Communication shapes how information flows, is interpreted, and acted upon across organizational boundaries, through shared meanings (threat taxonomies, risk ratings, incident classifications) and the interpretative schemes that let analysts make sense of incoming data.

**MLAA in T2** is most clearly illustrated through multi-level information flows across all three cases. The Australian bank operates the most sophisticated cross-level T2 structure: a formally designed three-tier threat intelligence model (tactical, operational, and strategic intelligence), delivered through a dual-bridge architecture (operational bridge for working-level coordination, management bridge for executive-level briefing). The Threat Intelligence Manager describes: *"Operational intelligence is basically we've been doing hunts and things like that. And we're monitoring the broader threat landscape and we want to pass intelligence to the operations team to uplift their existing platforms."* Indonesia's CIO describes sector-level T2: *"if someone from NAB... basically says hey guys, we found this in our environment, that's more relevant to us than your Mandiant telling us."* Malaysia uses FinTIP formally and informal inter-bank WhatsApp contacts.

**TAS and MASP in T2:** TAS combines stable protocols with adaptive intelligence content; MASP operates through machine-to-machine communication (automated alert sharing, API-driven IOC feeds, cross-tool correlation) at speeds unattainable by humans. A notable cross-case finding concerns informal channels. War-room coordination in Indonesia and Malaysia runs on Microsoft Teams. WhatsApp is used alongside this as a supplementary, informal cross-boundary channel, alongside email, culturally embedded but insecure. Indonesia's CIO notes: *"WhatsApp - a lot of things are done through WhatsApp, and yeah I know, it's not secure."* The fastest channel is thus the least secure; the Australian bank's formal dual-bridge architecture is the contrast case.

### 4.5.3 Triad 3: Organizational Composition and Power Dynamics (Authority Structure)

Authority structures determine who has decision rights, allocates resources, and coordinates organizational responses, through formal and informal hierarchies (domination) and the organizational mechanisms through which power is exercised (facility).

**TAS in T3:** The cross-case comparison suggests T3 may be among the most consequential triads for understanding resilience variation. The Australian bank's restructure represents a slow-cycle core structural innovation with fast-cycle enabling implications: by removing middle management, the IR Team Leader gained direct access to all other organizational domains without approval bottlenecks. *"The* *restructure is about taking out middle management... I am empowered to go directly to the domain leads of all the other domains and engage them at any point in time."* Indonesia's CISO holds dual authority (CSIRT head and security governance lead), enabling rapid command authority. Malaysia's MBRC structure provides clear governance but the Tech Risk Lead identifies a cultural T3 barrier: *"It's a power thing, so a superior has control over a junior... they cannot jump over their direct report."* This hierarchy constrains T2 communication flows, creating a cross-triad tension visible only through STARC.

**MLAA and MASP in T3:** power is distributed across levels, while MASP adds algorithmic enforcement (blocking, quarantining, denying access) under human override, producing hybrid power structures that require meta-governance.

### 4.5.4 Triad 4: Policy, Procedures, and Compliance (Legitimation Framework)

Legitimation structures underpin cybersecurity practice through policies, standards, and regulatory requirements - here Australia's APRA/ASIC regime, Indonesia's OJK/Bank Indonesia/BSSN/PDP Law 2022, and Malaysia's Bank Negara RMIT mandate.

**TAS in T4:** Core stability versus peripheral adaptation is most visible here. In all three cases, regulatory requirements (APRA standards, OJK SE 29, Bank Negara RMIT) and fundamental security principles form slow-changing core legitimation structures, as Indonesia's Team Leader notes: *"in essence, we must have an SOP for every bank... When a cyber incident occurs, we can do what is called mitigation or perhaps recovery."* Indonesia faces the most complex landscape, harmonizing BI, NIST, and BSSN frameworks with biennial SOP review, while Malaysia's 2019 RMIT requirement directly created the CISO function.

**MLAA and MASP in T4:** MLAA shows legitimation operating across levels, with regulatory mandates constraining tactical flexibility; Indonesia's one-hour reporting window (*"the regulator had to take an hour to report an incident"*) imposes fast-cycle legitimation pressure. MASP appears through automated compliance monitoring, audit logging, and policy-violation detection alongside human exception handling.

### 4.5.5 Triad 5: Technology and Human Resources (Material-Agential Facilities)

Technology and human resources form the material and cognitive foundation enabling effective cybersecurity operations; the three cases reveal dramatically different T5 maturity profiles.

**MASP in T5 (Primary Emphasis):** Autonomous agential capacity varies markedly across cases. The Australian bank exhibits the highest - full automated detection, containment, and forensics for known malware, with the IR Team Leader noting *"that particular scenario didn't require our analysts to intervene."* Indonesia exhibits high capacity, with a dedicated detection-engineering team building custom rules: *"we have a team to develop the rules... Every day as we receive we do research to make a new detections."* Malaysia exhibits developing capacity, relying on an MSSP for L1–L3 triage and open-source intelligence while a commercial platform is procured, its SIEM *"still requires some fine tuning."*

**Human T5 resources** vary as well. The Australian bank runs a fully insourced 24/7 SOC with formal experience tiers (L1: 1–2 years; L2: 2–3; L3: 5–10); Indonesia runs a hybrid model (outsourced shift analysts under internal managers, with internal responders bridging technical and organizational response); Malaysia relies heavily on MSSP vendors (*"We are relying quite heavily on them"*). These differences propagate into T1 and T2 maturity, from the Australian bank's dual playbook architecture to Indonesia's KPI-driven capability development.

**MLAA and TAS in T5:** MLAA frames resources across levels; TAS distinguishes stable foundations (core platforms, capabilities) from adaptive elements (rules, configurations) enabling rapid adjustment.

## 4.6 Cross-Triad Integration and Model Dynamics

The triads operate simultaneously but with varying emphasis: T1 and T5 dominate routine operations, T1–T3 come to the foreground during incident response, and T4 enters adaptive mode during post-incident learning.

**Theoretical Proposition 6 (P6):** Organizational cyber resilience is enhanced when communication structures (T2) align with power structures (T3), enabling rapid decision escalation and coordinated response during incident contexts.

**Theoretical Proposition 7 (P7):** Organizations that develop second-order learning structures enabling rapid modification of first-order threat response structures (T4 meta-policies enabling T1–T5 adaptation) exhibit superior cyber resilience.

Table 2 summarizes the propositions and the cross-case evidence supporting each, together with suggested research designs for empirical testing.

*Table 2: STARC Testable Propositions, Cross-Case Evidence, and Empirical Research Agenda*

| P | Statement | Extension/ Triad | Cross-Case Evidence | Research Design |
|---|---|---|---|---|
| P1 | Multi-level agency alignment enhances resilience; misalignment creates vulnerabilities. | MLAA / All | Australia: ACSC/peer-bank community triggers proactive control deployment. Indonesia: BSSN/Perbanas community enables pre-attack detection. Malaysia: FinTIP intelligence sharing. | Multi-level longitudinal study tracking agency alignment across individual, organizational, sector, institutional levels. |
| P2 | Core structural stability + rapid peripheral adaptation → enhanced resilience. | TAS / T1,T4,T5 | Australia: stable CSIRT charter enabled 20-min automated crypto-miner response. Indonesia: SDLC gate + daily rule updates. Malaysia: BCM codes + IOC-driven scanning. | Comparative cases measuring time-to-adapt (peripheral) vs. durability (core); correlate with resilience outcomes. |
| P3 | Rigid structures and fluid structures both reduce resilience vs. bounded mutability. | TAS / T3,T4,T5 | Indonesia: pre-2018 fluid (scramble) vs. post-2018 bounded. Malaysia: developing (playbooks in draft). Australia: mature bounded mutability (restructure + peripheral flexibility). | Multi-case study: rigid vs. fluid vs. bounded mutable organizations. Test non-linear relationship with resilience outcomes. |
| P4 | Temporal differentiation across triads (fast tactical, slow strategic) enhances resilience. | TAS / All | Australia: 20-min automated cycles vs. multi-year restructure. Indonesia: 15-min MTTD/30-min MTTR vs. biennial SOP review. Malaysia: 1-2 hr MBRC vs. annual policy review. | Longitudinal process study tracking cycle times across triads; correlate temporal differentiation patterns with resilience. |
| P5 | Meta-governance of algorithmic-human boundaries enhances resilience. | MASP / T3,T5 | Australia: CISO decides takedowns; incident manager coordinates. Indonesia: CISO recommends, business decides (gap acknowledged). Malaysia: MBRC decides network-wide actions. | Mixed-methods documenting boundary policies. Measure incidents from boundary failures; correlate with resilience outcomes. |
| P6 | Communication and power structure alignment enables rapid escalation and coordinated response. | MLAA+TAS / T2,T3 | Australia: dual-bridge aligns T2-T3; restructure direct access. Indonesia: CISO command + tiered WhatsApp. Malaysia: RACI matrix, but hierarchy constrains escalation. | Comparative incident analysis: communication path length, decision latency, coordination effectiveness across organizations. |
| P7 | Second-order learning structures → superior first-order adaptation and resilience. | TAS+MLAA / T4,All | Australia: formal PIR + purple teams (but L1 gap). Indonesia: KPI drills + lesson-learned to rule updates. Malaysia: PIR largely absent (negative case). | Longitudinal comparative: organizations with/without second-order structures. Measure adaptation speed and effectiveness. |

## 5. What STARC Illuminates: Theoretical Comparison

The following comparison is a heuristic illustration, not a confirmatory empirical test. Applying STARC alongside prior lenses to the same material identifies distinctions it makes visible - not to argue prior theories are inadequate, but to show its apparatus foregrounds phenomena single-level, single-temporal, or anthropocentric frameworks do not. STARC is thus integrative and complementary, extending rather than replacing the theories in Section 2. Equally important, the three-case material is diverse instantiation, not a ranking: each case is included for the region of the framework its configuration makes observable, and the cases are not scored against one another on a common scale of maturity, clarity, or resilience. What differs is which mechanisms each configuration brings into view, and that difference carries the analysis.

### 5.1 Insight 1: Multi-Level Agency Alignment as Resilience Mechanism

Ahmad et al. (2021) and Kotsias et al. (2022) analyzed the Australian bank's situational awareness and organizational learning, respectively, but neither foregrounded the recursive connections between individual awareness and the organizational, sector, and institutional structures that shape it.

**STARC Lens:** Through STARC, sector-level intelligence (MLAA) flows to organizational platforms (T5) and shapes analyst schemes (T1-T2), making awareness actionable before incidents occur. The recursive dimension is illustrated by the bank's Threat Intelligence Manager, who describes sector-level agency through peer-validated intelligence: *"If someone from NAB or CBA... says we found this in our environment, that's more relevant to us than your Mandiant telling us... threat intelligence sharing through people in your region, in your sector... it's probably critical for us."* Cross-case comparison extends this: Indonesia's peer-bank intelligence (the peer institution attack) triggers a proactive posture and Malaysia's FinTIP drives threat hunting - consistent with P1 (multi-level agency alignment enhances resilience). Malaysia's configuration, with detection largely outsourced and sector coordination through informal channels, foregrounds a different facet of P1 - where multi-level agency dependencies sit when a bank does not hold threat-attribution capability in-house - a distinct region of the proposition rather than a less complete version of the same alignment, with sector-level agency accessed through external providers.

### 5.2 Insight 2: Bounded Mutability Through Temporal Differentiation

**STARC Lens:** Where Situational Awareness and Organizational Learning theories offer limited apparatus for simultaneous stability and rapid adaptation, the TAS lens shows resilience emerging not from learning speed alone but from appropriate temporal differentiation - fast where speed matters, slow where stability matters. The Australian configuration renders bounded mutability in realized form, in the IR Team Leader's account of an automated cloud-container response: *"09:02 our detection rule kicked in... 09:05 the environment was contained and locked down... By 09:20 we had completely rebuilt the environment."* The stable core (detection rules, automation, CSIRT charter) enabled this 20-minute response without governance renegotiation. Indonesia's post-2018 reconstruction illustrates P3 from the negative side: pre-2018 fluid structure (scramble mode) produced a major breach; post-2018 bounded structure is associated with markedly improved resilience. The Malaysian configuration, lacking a settled cybersecurity-specific core, renders a different facet: peripheral adaptation proceeding while core boundaries are still drafted - the condition P3 identifies as risking chaotic rather than bounded adaptation, as the Tech Risk Lead notes the cyber incident playbook remains in draft. Each configuration instantiates a different region of P2 and P3 - realized, reconstructing, and forming - consistent with, though not confirmatory of, the propositions.

### 5.3 Insight 3: Algorithmic-Human Authority Boundary Management

**STARC Lens:** Where prior analyses centered on human cognition without addressing where algorithmic-human authority boundaries lie, MASP brings that challenge into view. Across the three cases it foregrounds three meta-governance questions rather than positions on a single clarity scale. The Australian bank manages the boundary explicitly: automated systems handle known-threat pattern matching at scale, while human authority handles service-impact decisions requiring contextual judgment. The CISO's authority over the 48-hour service takedown decision - *"the CISO ultimately made the decision and we were then able to proceed"* - exemplifies this. Indonesia's acknowledged meta-governance ambiguity (*"the decision maker is not yet clear"*) shows the same boundary before it settles - an authority relation being constituted rather than exercised, which P5 addresses at the point of construction, not a lower score on a clarity scale.

Malaysia's configuration extends the boundary to the MSSP (BCM advises, MBRC decides), so the question becomes where authority runs when triage is external. Each configuration thus makes a different meta-governance question visible - who holds the boundary, how it is constituted before settling, and where it runs when detection is outsourced - the range P5 captures. It should be noted that the precise degree of machine learning autonomy in these systems cannot be determined from interview data; MASP applies across the rule-based-to-ML spectrum, since the governance challenge is present regardless of technical detail.

### 5.4 Insight 4: Power-Communication Structural Alignment

**STARC Lens:** Prior analyses left the alignment of communication (T2) and power (T3) unexamined; through the T2-T3 lens, the Australian bank's restructure reads as strategic T2-T3 alignment - direct communication channels carrying direct authority, so information translates to action without delay - not merely efficiency optimization. The Malaysian configuration instantiates the misalignment cost: hierarchy constraining junior-to-senior escalation creates T3 barriers to T2 speed. Indonesia's WhatsApp dependency shows a different misalignment: informal channels give T2 speed without T3 authority clarity, producing fast communication but ambiguous action authority. These contrasts are consistent with P6.

### 5.5 Insight 5: Second-Order Learning as a Stratified Capability

Organizational Learning Theory addresses how organizations learn from incidents but not the structural conditions under which learning is institutionalized across levels rather than confined to senior tiers.

**STARC Lens:** Through STARC, post-incident learning (P7) is a T4 meta-policy enabling T1–T5 adaptation, revealed by the three-case comparison as stratified rather than uniform. The Australian configuration concentrates formal second-order learning at senior tiers - post-incident reviews, purple-team exercises, tabletop drills - yet even here a revealing contradiction surfaces. The IR Team Leader describes a formal review process: *"we then go through... a post incident review... root cause analysis of what had happened, why it happened and take away the learnings."* The Level 1 SOC Analyst reports the opposite: *"Once the ticket is closed and the incident is over, is there a formal PIR?... So far it hasn't."* It is retained as a partial (negative) case: learning reaches senior tiers but not consistently Level 1, so P7 is realized unevenly even where these mechanisms are most fully developed. The SOC Leader corroborates this, noting learning below formal reviews is not reliably retained: *"We're probably not good at preserving it but we're trying to... someone might have that on their own machine or something and may lose it."* The Indonesian configuration shows an institutionalized learning-to-detection loop, with lessons-learned feeding directly into detection rules: *"recommendations will appear in the lesson learned, then we... apply those parameters in the tool."* Malaysia is the clearest negative case, with the VP of Information Security noting: *"after the incident is over, you do not come back and revisit it. So there's no post review process."* Read together, these are three learning configurations, not points on one scale: an institutionalized learning-to-detection loop, formal learning concentrated at senior tiers, and learning not yet routinely revisited after incidents. The apparent tension - that the most institutionalized loop is not the most resilient overall - dissolves rather than contradicts P7: learning configuration and resilience are separable facets, the latter also shaped by resourcing and outsourcing. The variation is consistent with P7, showing the value of treating learning as structural rather than purely cognitive.

## 6. Theoretical and Practical Contributions

### 6.1 Contributions to Cybersecurity Research

STARC makes five contributions to cybersecurity research, each addressing an unresolved question in the literature.

First, STARC explains why resilience varies across organizations that the 'people, process, technology' framework would treat as comparable on paper: all three run SIEM platforms and comply with sector mandates, yet differ fundamentally in resourcing and outsourcing structure. STARC accounts for this not through a faster-versus-slower comparison but through four structural dimensions (cross-triad alignment, multi-level agency coordination, temporal differentiation, and meta-governance clarity) that explain why differently resourced and outsourced configurations produce qualitatively different response processes, moving beyond control inventories.

Second, STARC reveals post-incident learning as stratified rather than uniform. Even at the Australian bank, formal root-cause review feeds governance adaptation while the Level 1 analyst reports none reaching the operational tier; Indonesia's loop feeds lessons into detection rules, while Malaysia's is largely absent (Section 5.5). Effectiveness depends not on formal existence but on the T4-to-T1 pathways to operational tiers, which managers should assess directly.

Third, STARC identifies an incident communication paradox: alongside the formal, Teams-based war-room channel used for primary coordination, both Indonesian and Malaysian staff also rely on WhatsApp for supplementary cross-boundary contact, outside audit trails and retention controls, whereas the Australian bank's dual-bridge architecture keeps equivalent T2 speed entirely within T4-compliant governance. The misalignment carries regulatory exposure (Indonesia's 2022 Personal Data Protection Law, Malaysia's data-governance obligations) and forensic gaps that may compromise investigation. Communication architecture should be deliberate T2-T4 alignment, not left to evolve informally.

Fourth, STARC explains how MSSP dependency interrupts the recursive learning sustaining resilience: as Malaysia's L1-L3 model (Section 4.5.1) shows, externally performed triage means tacit knowledge accrues to the vendor, leaving T1 agency disconnected from the organizational agency it should inform, illustrated by the Head of Surveillance's account of heavy day-to-day reliance on the MSSP for triage. Organizations relying on MSSPs should design knowledge-transfer and hybrid staffing that preserve recursive feedback, not treat engagement as simple outsourcing.

Fifth, STARC grounds detection engineering theoretically, a discipline otherwise lacking foundation beyond operational best practice. Positioned between T1 (operational foundation) and T5 (the facilities that constitute detection capability), detection rules are the paradigmatic peripheral element adapting within stable core boundaries - illustrated by Indonesia's daily custom-rule development within a stable SDLC-governed architecture and the Australian bank's hours-to-deployment WAF/IPS rules (Section 4.2.1). This connects detection engineering to organizational-design questions: how teams sit relative to the SOC, how outputs are governed, and how work feeds into learning via the T5-to-T4 loop.

Finally, MASP's meta-governance apparatus addresses the SOAR governance challenge - which actions execute automatically versus require human authorization, the boundary P5 proposes as a resilience determinant (Section 5.3) - letting practitioners define automation thresholds and human-in-the-loop triggers as capabilities evolve.

### 6.2 Contributions to Structuration Theory

STARC extends Structuration Theory in ways applicable beyond cybersecurity. TAS theorizes structures operating at different temporal scales and external adversarial agency as a continuous exogenous pressure requiring active structural maintenance; unlike organizational ambidexterity theory, it specifies both the mechanism (bounded mutability) and the driver (adversarial intelligence).

MASP extends ST's anthropocentric foundations by theorizing algorithmic agency across a spectrum from rule-based to machine-learning-driven autonomy, requiring meta-governance structures managing human-algorithm boundaries. This is distinct from sociomateriality theory's analytical inseparability stance: MASP maintains analytical distinction between human and algorithmic agency as the object of governance interest. Unlike ANT's actant framework (Latour 2005; Callon 1986), which dissolves the structure/agency binary, MASP extends ST's duality structure to accommodate algorithmic properties while retaining the analytical purchase that the structure/agency distinction provides.

MLAA provides theoretical apparatus for analyzing recursive cross-level agency mechanisms in adversarial contexts, extending Stones' (2005) strong structuration to multi-organizational settings. Unlike multi-level analysis in IS research, which provides methodological tools for statistical aggregation, MLAA theorizes the recursive mechanisms through which agency at one level constitutes agency at other levels.

STARC's propositions generate measurable variables for empirical research: multi-level agency alignment (P1) can be operationalized as the latency between sector-level threat intelligence receipt and organizational control deployment; bounded mutability (P2–P3) can be measured as the ratio of core-to-peripheral structural change rates; meta-governance effectiveness (P5) can be measured through the rate of algorithm-attributed false positives requiring human review.

### 6.3 Practical Implications

As a diagnostic tool, STARC lets organizations assess resilience across triad strength (T1–T5), cross-triad integration, multi-level agency coordination, temporal differentiation, and algorithmic-human boundary management - a more comprehensive vocabulary than technical control inventories alone.

Bounded mutability offers a design principle: practitioners can deliberately separate what must stay stable from what must adapt rapidly, as the Australian and Indonesian cases illustrate (Sections 4.5.3, 5.2).

As AI adoption accelerates, MASP highlights explicit meta-governance of algorithmic-human boundaries as a critical strategic challenge. The Indonesian case - where decision authority was acknowledged as unclear - illustrates the risk of leaving this boundary implicit. Organizations should invest in policy instruments defining which alert categories trigger automated action without human review, which decisions always require human authority regardless of algorithmic confidence, and how those boundaries shift as system capabilities evolve.

STARC highlights cybersecurity as an inherently multi-organizational phenomenon. Organizations should invest in sector-level communication structures (ISACs, threat intelligence sharing, BSSN/ACSC coordination) as direct enablers of organizational resilience through MLAA. The contrast between the Australian bank's rich sector community (enabling advance warning of geographical TTP lag) and Malaysia's developing FinTIP participation illustrates the resilience value of this investment.

## 7. Conclusion

STARC addresses the need for multi-level theoretical understanding of cybersecurity practice. Through its three innovations - MLAA, TAS, and MASP - it foregrounds cross-level coordination, offers a structuration-theoretic account of simultaneous stability and adaptation, and theorizes the governance of hybrid human-algorithmic decision-making. The illustrative application across three banks in Australia, Indonesia, and Malaysia, spanning 20 interviews, provides cross-case variation consistent with the seven propositions, while acknowledging this grounding is designed for subsequent empirical testing rather than confirmatory validation. The model's implications extend beyond cybersecurity to broader Information Systems research on adversarial contexts and autonomous agency, illustrating how foundational sociological theory can be extended to address contemporary socio-technical challenges.

### 7.1 Limitations and Future Research

The study has several limitations. The illustrative application draws on three financial-sector organizations, so the grounding may not capture STARC's full range of manifestations across other sectors, sizes, or regulatory environments. The three-case dataset, while richer than prior single-organization work, remains illustrative rather than confirmatory. The degree of machine-learning autonomy in the systems described - particularly the Australian bank's Advanced Cyber Analytics platform - cannot be determined from interview data; future work should gather technical specifications alongside interviews. Distinguishing rule-based from machine-learning-driven systems at the technical specification level would also enable finer-grained testing of whether MASP's governance implications differ systematically across the autonomy spectrum. Future research should test the seven propositions through multi-level longitudinal studies (P1, P4), comparative case studies of bounded mutability across sectors (P2, P3), and mixed-methods designs documenting meta-governance policies (P5).